# Taking the reaction-diffusion master equation to the microscopic limit

Paul Sjöberg[1], Otto G Berg[2] and Johan Elf[1]

[1] Linneaus Center for Bioinfomatics in Department of cell and molecular biology, Uppsala University, Husarg. 3, SE-75124 Uppsala, Sweden [2] Department of evolution, genomics and systematics, Uppsala University, Norbyv. 18C, SE-75236 Uppsala, Sweden

## Abstract

The reaction-diffusion master equation (RDME) is commonly used to model processes where both the spatial and stochastic nature of chemical reactions need to be considered. We show that the RDME in many cases is inconsistent with a microscopic description of diffusion limited chemical reactions and that this will result in unphysical results. We describe how the inconsistency can be reconciled if the association and dissociation rates used in the RDME are derived from the underlying microscopic description. These rate constants will however necessarily depend on the spatial discretization. At fine spatial resolution the rates approach the microscopic rate constants defined at the reaction radius. At low resolution the rates converge to the macroscopic diffusion limited rate constants in 3D, whereas there is no limiting value in 2D. Our results make it possible to develop spatially discretized reaction-diffusion models that correspond to a well-defined microscopic description. We show that this is critical for a correct description of 2D systems and systems that require high spatial resolution in 3D.

# Introduction

Quantitative analysis of intercellular reaction networks will in many cases need to consider both the spatial and stochastic aspects of chemical processes. Spatial, because diffusion is not sufficiently fast to make the system well-stirred between individual reaction events. Stochastic, because the number of reactants within diffusion range commonly is low, such that the probabilistic and non-linear nature of chemistry invalidates mean-field descriptions. In recent years a number of strategies to model and simulate stochastic reaction-diffusion systems have been suggested (ChemCell, Smoldyn, GRFD, MesoRD, SmartCell, MCell etc.). These can be traced back to the two different basic theoretical frameworks for describing chemical reaction in dilute solutions; the spatially and temporally continuous Smoluchowski framework (von Smoluchowski, 1917) and the spatially discretized reaction-diffusion (or multivariate) master equation (RDME, (Nicolis and Prigogine, 1977) (Gardiner et al., 1976)). Including its extension to non diffusion limited (Collins and Kimball, 1949; Noyes, 1961) and reversible reactions (Berg, 1978) the former continuous description is clearly more fundamental, whereas the coarse grained RDME is better suited for mathematical analysis involving more than two molecules (Lee and Cardy, 1995) and for large scale simulation (Fange and Elf, 2006).

In RDME, space is divided into subvolumes. It has been suggested that these should be smaller than the mean free path between reactions, such that subvolumes can be considered well-stirred (Baras and Mansour, 1996). They should at the same time be larger than the mean free path between collisions with solvent molecules, so that movement can be considered diffusive. The more demanding condition on the lower boundary is however that subvolumes need to be sufficiently large for molecules to lose correlation in the subvolume between reactions (Baras and Mansour, 1996; Elf and Ehrenberg, 2004). The latter constraint is actually too restrictive and would for instance not be possible to satisfy in 2D, as will be shown in this letter.

In the RDME the state of the system is defined as the number of molecules of each species in each subvolume. The state changes when chemical reactions occur in a subvolume or when a molecule diffuses between subvolumes. These events are considered elementary in the sense that they have a constant probability to occur each infinitesimal time interval. Furthermore, the probability for a reaction or diffusion event only depends on the instantaneous local concentration in the subvolume. For example the probability that the first order event $A \xrightarrow{k} \varnothing$ occurs during $\delta t$ is $\delta t k \Omega a$, where $a$ is the concentration of A in the subvolume Ω is the volume of the subvolume. Similarly the probability that the association event $A + B \xrightarrow{k} C$ occurs is $\delta t k \Omega a b$. Diffusion events are considered first order reactions such that the probability that an A molecule jumps from one subvolume to a neighbour during $\delta t$ is $\delta t k_{diff} \Omega a$, where the jump rate $k_{diff}$ is chosen to satisfy the diffusion equation. For example $k_{diff} = D/l^2$ for cubic subvolumes with side length $l$. Taken together, these events define a RDME that describes how the probabilities change over the state space as a function of time.

Because of its relative simplicity the RDME framework has been commonly used both in physics, chemistry and biology over the decades. However, with the recent explosion of computational systems biology there has been a growing interest in how RDME is related to more detailed descriptions (Isaacson, 2008; Erban and Chapman, 2009). Two important remaining issues are how RDME relates to reversible reactions in the Smoluchowski description at the microscopic level, and how the spatial dimension influences the RDME model. In this letter we answer these questions. We will start from the microscopic model for a reversible interaction between two molecules in the Smoluchoswki framework with the microscopic boundary condition from Collins and Kimball (1949). We introduce a spatial discretization of the partial differential equation (PDE) that can be directly interpreted as a RDME, where the association and dissociation rates in the RDME are identified as boundary conditions for the PDE. Next we derive a mathematical model for how the discretized boundary conditions depend on the spatial discretization as well as the microscopic rate constants. Finally we use the scale dependent rate constants to demonstrate that it is possible to make a RDME involving many molecules that is consistent with the microscopic description.

## Methods and Results

The spatial aspects of chemical reactions are important for association and dissociation reactions rates since they depend on correlation between two molecules (Noyes, 1961; Berg, 1978). Irreversible zeroth and first order events do however not have any spatial dependence. We will therefore focus on the reversible chemical complex formation that can be represented by the following scheme,

$$A + B \underset{k_d}{\overset{k_a}{\rightleftarrows}} C, \qquad (1)$$

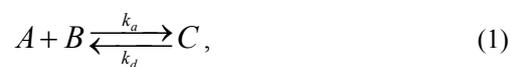

where $k_a$ is the association rate constant and $k_d$ is the dissociation rate constant. These macroscopic rate constants $k_a$ and $k_d$ are defined in a volume that is much larger than the molecules themselves. These rates generally depend on how fast the molecules diffuse, their reaction radius and how fast they react when they meet. The reaction radius is the distance at which molecules associate and dissociate microscopically. For example, in 3D the macroscopic association rate is given by (Collins and Kimball, 1949)

$$k_a = \frac{4\pi\rho D k}{4\pi\rho D + k}, \qquad (2)$$

where $k$ is the microscopic association rate, $\rho$ is the reaction radius, and $D$ is the sum of diffusion rate constants for the two reactants (Noyes, 1961).. The microscopic association rate $k$ is defined such that $\delta t k b_\rho$ is the probability that an A molecule will bind during time $\delta t$ if there is a concentration $b_\rho$ of B at the reaction radius. In the limit of fast diffusion, i.e. $4\pi\rho D \gg k$, we obtain $k_a = k$. On the other hand, the diffusion limited association rate is $k_a = 4\pi\rho D$.

Similarly, the macroscopic dissociation rate in 3D is

$$k_d = \frac{4\pi\rho D \gamma}{4\pi\rho D + k}, \qquad (3)$$

where $\gamma$ is the microscopic dissociation rate. Such a dissociation event positions the molecule at a distance of the reaction radius. It may seem strange the macroscopic dissociation rate constant depends on the rate of diffusion and the microscopic association rate constant $k$. This is however necessarily the case because macroscopic dissociation is a competition between immediate reassociation and separation by diffusion (Berg, 1978). On average the molecules will bind back $(4\pi\rho D + k)/(4\pi\rho D)$ times before they lose spatial correlation. The equilibrium constant $K_d = k/\gamma = k_a/k_d$ does however not depend on the diffusion constant.

The relations (2) and (3) are derived from a microscopic model based on the Smoluchowski framework extended to reversible and non-diffusion limited reactions (below). This approach does not work in 2D where the macroscopic rate constants are not well-defined. However, to tie in directly with the RDME framework, it is more appropriate to consider the mean times for association in a finite region, which are well-defined in both 2D and 3D. In 3D, the mean-time approach gives the same macroscopic association rate constant as derived from the Smoluchowski approach. In 2D, however, increasing the size of the region leads to an ever decreasing association rate constant.

We will now use this framework as the microscopic reference and study a two-particle system. Without loss of generality, one of the particles defines the center of a spherical coordinate system. The other particle, the ligand, is freely diffusing with a diffusion rate constant that is the sum of the two particles' diffusion rate constants in a common reference system. The distance between the molecules' centers of mass is denoted $r$. Let $p(r,t)$ be the probability density for the ligand to remain unbound and separated from the target by $r$ at time $t$ and $p_b$ the probability for a bound state at time $t$. The time evolution of the system is then determined by

$$\begin{cases} \dfrac{\partial p(r,t)}{\partial t} = D\dfrac{1}{r^{\omega-1}}\dfrac{\partial}{\partial r}\left(r^{\omega-1}\dfrac{\partial p(r,t)}{\partial r}\right) \\ \dfrac{dp_b(t)}{dt} = kp(\rho,t) - \gamma p_b(t), \end{cases} \quad (4)$$

where $k$ is the microscopic association rate constant, $\gamma$ is the microscopic dissociation rate constant, $D$ is the diffusion rate constant and $\omega = 3$ in 3D and $\omega = 2$ in 2D. The microscopic rate constants are defined by the boundary condition of the diffusion equation at the interface $r = \rho$

$$D\varepsilon\left.\dfrac{\partial p(r,t)}{\partial r}\right|_{r=\rho} = -kp(\rho,t) + \gamma p_b(t) \quad (5)$$

where $\varepsilon = 4\pi\rho^2$ in 3D and $\varepsilon = 2\pi\rho$ in 2D. At the reflective boundary $r = R \gg \rho$,

$$\left.\dfrac{\partial p(r,t)}{\partial r}\right|_{r=R} = 0. \quad (6)$$

The red curves in Fig 1 show the time evolution for the probability of being in the bound state $p_b(t)$ assuming that the particles are bound from the start, i.e. $p_b(0) = 1$ and $p(r,0) = 0$. We will now use the system described by Eqs. 4-6 as the microscopic reference and study a two-particle system in a spherical reaction volume with radius R.

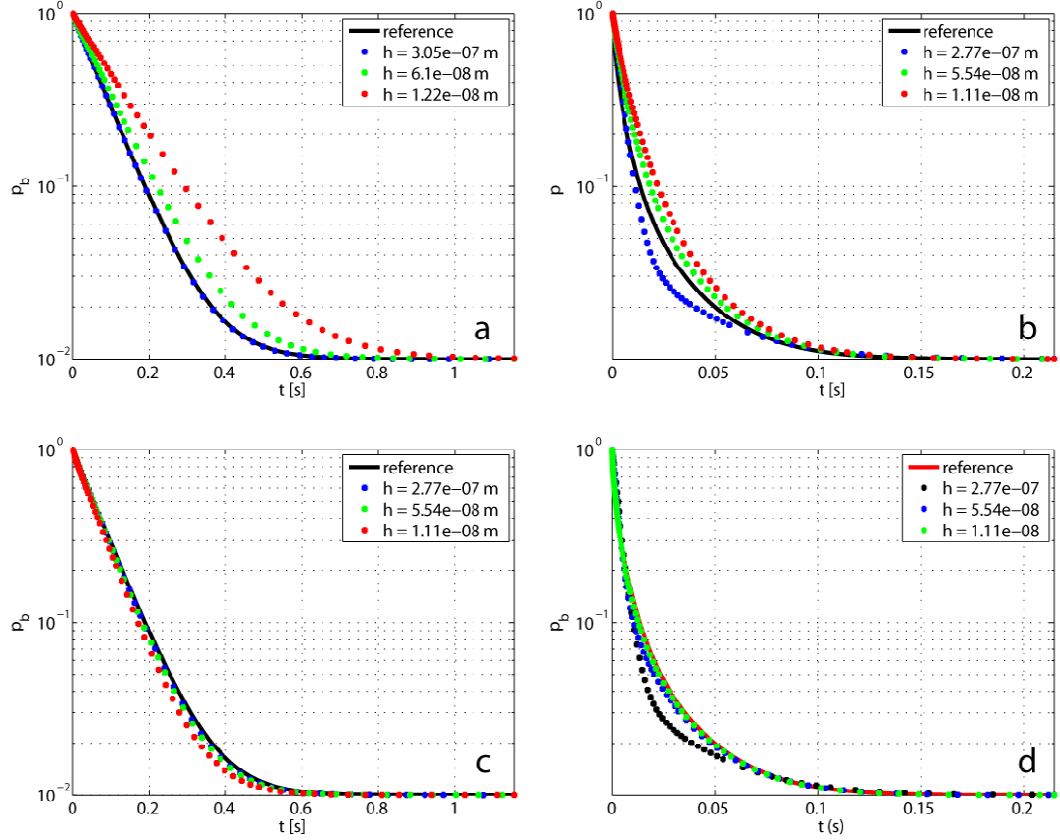

**Figure 1. Approaching equilibrium binding for two molecules at different RDME discretizations (h inset).** Diffusion-limited reactions with $D=1\mu m^2/s$, $\rho=10nm$ and association rate constants chosen such that the degree of diffusion control $\alpha=100$ and dissociation rate constants chosen the probability of the bound state at equilibrium is 0.01 in 3D. Using total volume $V=1\mu m^3$ (a, c), and 2D, with total area $V=1\mu m^2$ (b, d) we have (a) macroscopic rate constants $k_a == 1.31\ 10^{-19} m^3 s^{-1}$ and $k_d= 13.0 s^{-1}$, (b) macroscopic rate constants $k_a= 2.33\ 10^{-12} m^2 s^{-1}$ and $k_d= 231 s^{-1}$, (c) rate constants according to Eq. (15) with $k=1.26\ 10^{-17} m^3 s^{-1}$ and $\gamma=1240\ s^{-1}$, and (d) rate constants according to (15) with $k=6.28\ 10^{-10} m^2 s^{-1}$ and $\gamma=62200\ s^{-1}$,

## The RDME

Our two particle system results in a simple reaction-diffusion master equation. The domain is discretized in $r$ by $n$ shell-shaped volumes using the step size $h$. It is only in the innermost shell reactions can occur. Let $p_j(t)$ be the probability for the ligand to be unbound in the volume where $r \in (\rho+(j-1)h, \rho+jh)$. Then master equation approximation of Equation (4) can be written

$$\begin{cases} \dfrac{dp_b(t)}{dt} = \dfrac{q_a}{V} p_1(t) - q_d(t) p_b(t) \\ \dfrac{dp_1(t)}{dt} = g_2 p_2 - f_1 p_1 + q_d(t) p_b(t) - \dfrac{q_a}{V} p_1(t) \\ \dfrac{dp_i(t)}{dt} = g_{i+1} p_{i+1} - (g_i + f_i) p_i + f_{i-1} p_{i-1} \\ \dfrac{dp_n(t)}{dt} = -g_n p_n + f_{n-1} p_{n-1}, \end{cases} \qquad (7)$$

where $i = 2,\ldots,n-1$, $f_j = D\omega \dfrac{r_j^{\omega-1}}{h(r_j^\omega - r_{j-1}^\omega)}$, $g_j = D\omega \dfrac{r_{j-1}^{\omega-1}}{h(r_j^\omega - r_{j-1}^\omega)}$, $r_j = \rho + jh$ and $V$ is the effective volume of the innermost subvolume, that is $V = (4\pi/3)((\rho+h)^3 - \rho^3)$ in 3D and $V = \pi((\rho+h)^2 - \rho^2)$ in 2D.

The question is which values should be used for the reaction rates $q_d$ and $q_a$. Conventionally one would use the macroscopic reaction rates, i.e. $q_a = k_a$ and $q_a = k_d$ (Haken, 1975; Gardiner et al., 1976; Lemarchand and Nicolis, 1976). However, in Fig 1 a and b we see how poorly the RDME describes the kinetics of the relaxation processes when we use the diffusion limited rate constants for different discretizations. It is unsatisfactory that the solution of the RDME depends on the arbitrary discretization in this way and also that the deviation from the correct curves gets more pronounced the finer the discretization. The reason for the poor behavior is that the reaction no longer is diffusion limited when the molecules end up in the same subvolume at fine discretization, where the diffusion aspect of the reaction is handled explicitly by the diffusive jumps. In the limit that we let $h \to 0$ Eq. (7) is in fact a simple numerical scheme to solve Eq. (4), in which case we obviously would use the microscopic rate constants, i.e. $q_a = k$ and $q_d = \gamma$. It appears that we need to adjust the rate constants used in the RDME such that the contribution of diffusion gets smaller at fine discretization.

In order to determine these effective or mesoscopic rate constants spanning the gap between the micro and macroscopic rates we will solve the continuous reaction-diffusion equation for the central subvolume $[\rho, \rho+h]$ under the constraints given by the RDME. Thus all movements in and out of the inner subvolume are accounted for by the jump probabilities between neighboring subvolumes at rates determined by the diffusion constant and geometry. The initial condition for the PDE is therefore a homogeneous probability density

$$p(r,0) = \dfrac{1}{V}, \qquad (8)$$

Where $V$ is the accessible volume of the innermost subvolume as defined above. We calculate the rate of the first association event in competition with the diffusive jump rate $f_1$ out of the subvolume. The diffusive jumps out of the domain are equally probable anywhere in the subvolume. The mean free time for a molecule in the volume, the residence time,

$$\tau_{res} = \frac{1 - p_{ass}}{f_1}, \tag{9}$$

where $p_{ass}$ is the probability for a molecule to associate to the target rather than jumping out. The effective rate constant is,

$$k_{eff} = \frac{p_{ass}}{\tau_{res}} = \frac{f_1}{1/p_{ass} - 1} \tag{10}$$

To determine $p_{ass}$, Eq. (4) with a homogeneous loss term representing diffusive jumps,

$$\frac{\partial p(r,t)}{\partial t} = D \frac{1}{r^{\omega-1}} \frac{\partial p(r,t)}{\partial r} - f_1 p(r,t) \tag{11}$$

is solved with a flow condition on the inner boundary

$$DA(\rho) \frac{\partial p(\rho,t)}{\partial r} = -kp(\rho,t) \tag{12}$$

and a reflecting outer boundary at $R=\rho+h$

$$DA(\rho) \frac{\partial p(R,t)}{\partial x} = 0, \tag{13}$$

where $A(r) = 2\pi r$ in 2D and $A(r) = 4\pi r^2$ in 3D. The association probability

$$p_{ass} = \int_0^\infty kp(\rho,t)dt \tag{14}$$

follows. The discretization-dependent solution can be expressed as

$$q_a(h) = \frac{k}{1 + \alpha G(\beta) - \alpha\beta(1-\beta)}, \tag{15}$$

where $\beta = \rho/(\rho+h)$ and $\alpha$ is the degree of diffusion control; in 2D, $\alpha = k/(2\pi D)$ and in 3D, $\alpha = k/(4\pi\rho D)$. $G(\beta)$ is a function determined only by geometry and is given by

$$G(\beta) = \frac{1}{\beta Q} \frac{I_1(Q)K_0(\beta Q) + K_1(Q)I_0(\beta Q)}{I_1(Q)K_1(\beta Q) - K_1(Q)I_1(\beta Q)} \quad \text{in 2D, and}$$

$$G(\beta) = \frac{(Q+1)e^{-2(1-\beta)Q} + Q - 1}{(Q+1)(1-\beta Q)e^{-2(1-\beta)Q} + (Q-1)(1+\beta Q)} \quad \text{in 3D,} \tag{16}$$

where $I$ and $K$ are modified Bessel functions of the first and second kind, and $Q$ also is a function of β

$$Q = \sqrt{f_1 R^2 / D} = \sqrt{\omega / \left[(1-\beta)(1-\beta^\omega)\right]}$$

In both cases the dissociation rate constant $q_d(h)$ is simultaneously determined by the constant ratio given by the equilibrium constant $q_a(h)/q_d(h) = k/\gamma = K$.

In Fig. 2 the effective mesoscopic rates $q_a(h)$ and $q_d(h)$ given by Eqs. (15) are plotted as function of $h$. As expected, the association rate constants approach the microscopic rate constants in the limit of fine discretization $(h \to 0)$. Subsequently $q_d$ approaches the microscopic dissociation rate $\gamma$ in the same limit. In 3D the mesoscopic association rate Eq. (15) converges to the diffusion limited macroscopic rate given by Eq. (2) when $h \gg \rho$. At the same time, the dissociation rate $q_d$ converges to the macroscopic reaction rate $k_d$. This dissociation event includes multiple microscopic reassociation events and ultimately loss of correlation between the dissociating molecules. In 2D however, there is no well-defined limiting value for $q_a$. Instead the mesoscopic association rate constant slowly decreases for larger discretizations. The reason for this is that the macroscopic rate constant in 2D is concentration dependent and the concentration decreases as the subvolume gets larger.

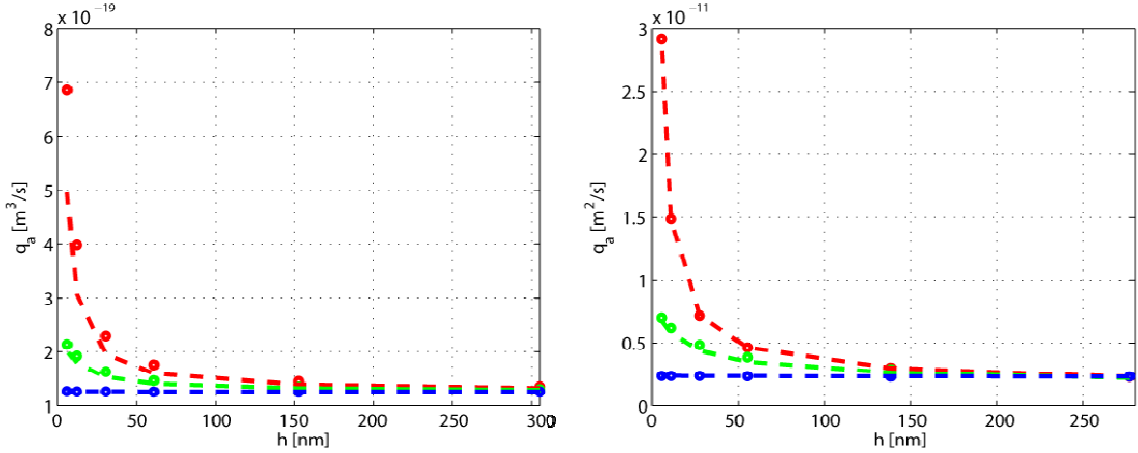

*Figure 2. Mesoscopic rate constants for different discretizations (h) and different degrees of diffusion control (α).* Results from analytical expression Eq. (15) (circles) are compared to numerically optimized result (dashed line) for α=100 (red), α=1 (green), α=0.01(blue) *Left* For the 3D system $V=1\mu m^3$, ρ=10nm. Both *D* and the microscopic rate (k) are changed such that the macroscopic rate constant is the same ($k_a$=1.27 $10^{-19}$ $m^3 s^{-1}$) for different degree of diffusion control . The microscopic dissociation rate γ is adapted such that the probability for being bound at equilibrium is 0.001.
*Right* For the 2D system $V=1\mu m^2$, ρ=10nm. Both *D* and the microscopic rate (k) are changed such that the rate constant is the same at h=2.5µm for different degree of diffusion control. The microscopic dissociation rate γ is adapted such that the probability for being bound at equilibrium is 0.001.

To test if the derived mesoscopic rates are appropriate also for intermediate discretizations we postulate that the probability for being bound or unbound should not depend on the discretization. In particular we demand that the relaxation rate $\lambda$, where

$$\lambda^{-1} = \int_0^\infty t \frac{dp_b(t)}{dt} dt \qquad (17)$$

should be independent of discretization and equal to the relaxation rate for the a highly resolved reference solution of Eq. (4). $\lambda$ will increase monotonically with $q_a$ (and $q_d = K_d \cdot q_a$). For each discretization there is a unique value for $q_a(h)$ such that the relaxation rate of the master equation is equal to that of the reference solution.

In Fig. 2 we see how this $q_a$ depends on the discretization for different degrees of diffusion control in 3D and 2D. These curves were obtained by minimizing the difference between the relaxation rate $\lambda$ (Eq.(17)) for the microscopic reference, Eq. (4), and the discretized RDME, Eq. (7). The numerically optimized mesoscopic rates are compared to those given by Eq. (15). We note that the agreement is excellent for different degrees of diffusion control, which implies that the analytical expression also leads to discretization independent relaxation rates. Fig. 2 also exemplifies that one and the same macroscopic rate constant corresponds to many different microscopic models at fine discretization. For this reason knowledge about the microscopic parameters is needed to make correct simulations at high spatial resolution.

Another way to test the validity of the mesoscopic rate constants Eq. (15) is to use them when solving the RDME for the two-molecule system, Eq. (7). In Fig. 1c and 1d these results are compared to the reference solution. It is clear that RDME evolutions using the mesoscopic rate constants are in far better agreement with the reference solution than the time evolutions with fixed rate constants seen in Figs. 1a and b. It is only for the coarsest discretization of the 2D system that it is not possible to accurately model the decay process with the RDME.

## 3. Examples

*1. Relaxation to equilibrium*

The mesoscopic reaction rate constants have been derived for pairs of molecules. In order to test if these rate constants can be used also when there are many molecules involved we have to rely on Monte Carlo simulation of the RDME. For this purpose we use the MesoRD software (Hattne et al., 2005) that implements an efficient Next Subvolume Method (Elf and Ehrenberg, 2004) for sampling trajectories from the RDME. When modeling a many particle system the RDME is defined in a Cartesian coordinate system that is common for all molecules. The volume is discretized into cubic subvolumes with side length $l$, that are taken to corresponds to the spherical volume including the reactants (i.e with radius $\rho + h$). Fig. 3 shows the relaxation kinetics of a system with 1000 complexes in 3D and 100 in 2D. For dashed lines the rate constants $q_a(\infty) = k_a$ and $q_d(\infty) = k_d$ are used and the relaxation kinetics is strongly dependent on the spatial discretization. The equilibrium point is however correct since $k_a / k_d = k / \gamma$. For the solid lines the mesoscopic rates $q_a(h)$ and $q_d(h)$ from Eq. (15) are used and the relaxation kinetics is practically identical for different at discretizations, except when the discretization is very close to the size of the actual molecules (2ρ=20nm). The reason for this deviation will be discussed in detail elsewhere. It can also be noted that at the most coarse level, the resolution is insufficient to describe the relaxation kinetics in the 2D system.

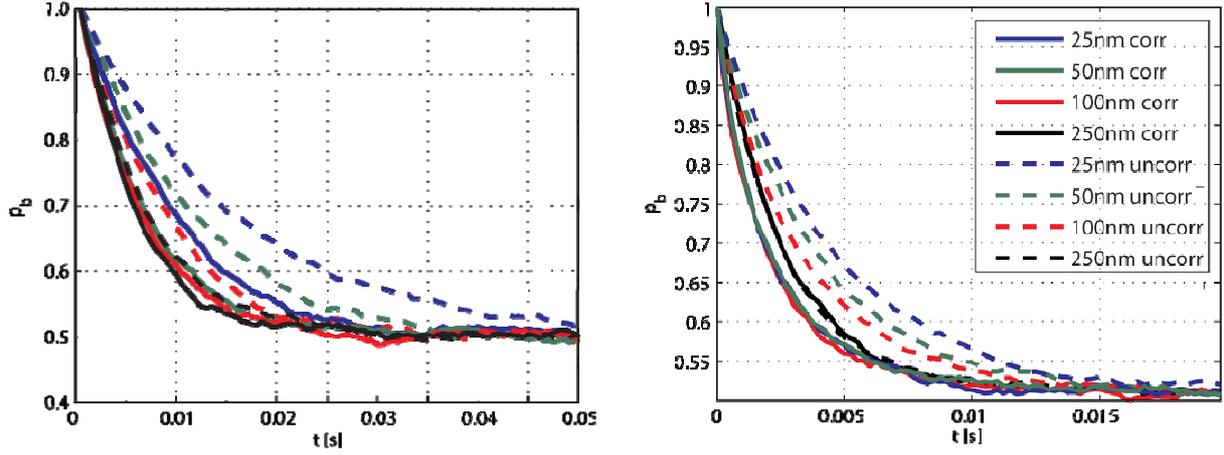

*Figure 3. Approaching equilibrium with RDME using different discretization (l, inset).* The reaction radius is 10 nm and the diffusion rate constant for each molecular species is 0.5 $\mu m^2 s^{-1}$. The association rate constant is chosen such that the degree of diffusion control $\alpha = 100$ and the dissociation rates are chosen such that the probability for an individual molecule to be bound in the complex C is 0.5 at equilibrium.
(Left) 3D system with $10^3$ molecules and total volume 1$\mu m^3$ averaged over 5 trajectories: Relaxation kinetics is plotted using rates according to Eq. (15) with k = 1.26 $10^{-17} m^3 s^{-1}$ and $\gamma$ = 6.28 $10^3$ s$^{-1}$ (solid lines) and for the fixed macroscopic rate constants $k_a$ = 1.24 $10^{-19} m^3 s^{-1}$ and $k_d$ = 62.2 s$^{-1}$ (dashed lines).
(Right) 2D system with $10^2$ molecules and total area of 1$\mu m^2$ averaged over 100 trajectories: Relaxation kinetics is plotted using rates according to Eq. (15) with k = 6.28 $10^{-10} m^2 s^{-1}$ and $\gamma$ = 3.14 $10^4$ s$^{-1}$ (solid lines) and for the fixed macroscopic rate constants $k_a$ = 3.46 $10^{-12} m^2 s^{-1}$ and $k_d$ = 173 s$^{-1}$ (dashed lines).

## 2. Non-Equilibrium Steady state

The spatial discretization of the RDME changes the kinetics of the system, but not the equilibrium point. However, in non-equilibrium situations, changes in kinetics of individual reactions typically lead to changes in the steady state. This implies that the spatial discretization of the RDME can change also the steady state of a system. To exemplify this extend the system Eq. (1) to

$$\begin{cases} A + B \underset{k_d}{\overset{k_a}{\rightleftarrows}} C \\ \varnothing \xrightarrow{k_1} C \\ A \xrightarrow{k_2} \varnothing \\ B \xrightarrow{k_2} \varnothing \end{cases} \quad (18)$$

The introduction of a zeroth order irreversible birth event and irreversible first or decay events are straight forward, since they do not depend on the spatial correlations between molecules. However, note that there is a big difference between the reversible first order dissociation event and a first order decay event. The former depends on a number of microscopic reassociation events and therefore also on the discretization whereas the latter, by definition, does not.

In Fig. 4 we see how the steady state copy number of C now depends on the spatial discretization unless we use the mesoscopic, scale-dependent, rate constants calculated from Eq.(15). In agreement with Fig 3a the corrected rate constants results a discretization dependant steady state, except when the discretization is close to the size of the reactants.

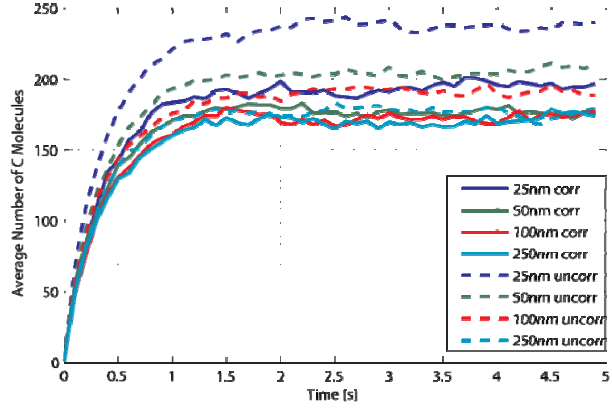

**Figure 4. The time evolution of the number of C molecules in the reaction system (Eq. 18) averaged over 20 realizations in 3D**. $k_1=1\mu Ms^{-1}$, $k_2=10s^{-1}$, $V=1\mu m^3$. The microscopic rate constants $k=2.51\ 10^{-17} m^3 s^{-1}$, $\rho=10nm$ and $D=1\mu m^2 s^{-1}$, such that the reversible reactions are diffusion limited ($\alpha=100$). $\gamma=2020 s^{-1}$, such that the macroscopic rate dissociation rate is equal to $2k_2$. The dashed lines are from simulations with the macroscopic rate constants $k_a= 2.49\ 10^{-19} m^3 s^{-1}$, $k_d=20s^{-1}$, and the solid lines are from simulations with the corrected mesoscopic rate constants.

## 4. Discussion

We find that the Reaction-diffusion Master Equation for a system with diffusion-limited reactions corresponds to different microscopic chemical systems at different discretization. This is because the reaction rates are kept constant, although in reality they shift from being diffusion-limited at coarse discretization to being reaction limited at fine discretization. We have shown how the rate constant can be calculated such that RDMEs at different discretizations are consistent with the same microscopic model. In particular, when the discretization length scale approaches the reaction radius $\rho$ information about the microscopic reaction rate constants $k$ and $\lambda$ as well as $\rho$ are needed to make a physically correct model. At coarse discretizations in 3D ($h \geq 10\rho$) many microscopic models converge and can be accurately described with certain diffusion limited rate constants, $k_a$ and $k_d$. In 2D systems there is no such convergence, and in theory the rate constants will have to be corrected on all length scales.

In the light of our results, the rules for how to choose the size of subvolumes need to be reevaluated. Concerning the upper limit, it was previously suggested that subvolumes should be smaller than the reaction free path (Baras and Mansour, 1996). This is obviously impossible for the diffusion limited reversible interactions that we have analyzed in this paper, since rebinding reactions occur also on the smallest length scales. Therefore the notion of a reaction free path needs to be redefined to exclude the microscopic rebinding events that are accounted for in the diffusion limited rate constants. When using the new mesoscopic rate constants the subvolumes should be smaller than the mean reaction free path for interactions between molecules that have not just have dissociated from each other.

Concerning the lower limit, it was previously suggested that the subvolumes need to be significantly larger than the reaction radii such that molecules can be fully dissociated within single subvolumes (Baras and Mansour, 1996; Elf and Ehrenberg, 2004). In 3D this would correspond to the case when the macroscopic rate constant has reached the macroscopic limit ($h \geq 10\rho$). In 2D it has not been possible to give a clear-cut lower limit of the subvolume size since the molecules do not loose correlation of previous interactions before they participate in other reactions. However, if subvolumes are this large they do not satisfy the constraint on the upper limit.

With the new mesoscopic rate constants the lower limit is now relieved and there is nothing that prevents decreasing the subvolumes to microscopic length scales comparable to the size of the molecules and the mean free paths between solvent interactions. This resolves the problem of modeling 2D systems since subvolume sizes can be chosen such that there is a clear separation between the previous dissociation event and following association to another molecule. It also makes it possible to simulate 3D systems where the mean free path between reactions is just a few molecule radii.

The new way to calculate reaction rates opens the possibility to make RDME simulation on unstructured grids (Engblom et al., 2009), where some subvolumes tend to become very small. It will also allow for the development of software solutions that seamlessly can change the spatial and temporal resolution within the same modeling framework, such that it is possible to find the optimal trade-off between accuracy and efficiency.